**Giant magnetic anisotropy of Pb atoms in 3d-based magnets**

Weiyi Xia, Cai-Zhuang Wang and Vladimir Antropov

Ames National Laboratory, U.S. Department of Energy, Ames, IA 50011, United States
Department of Physics and Astronomy, Iowa State University, Ames, IA 50011, United States

**Abstract**
Electronic structure analysis is performed to study the properties of several Pb-containing 3*d*-intermetallics. Our study reveals that binary metastable $Co_3Pb$ and $Fe_3Pb$ intermetallic compounds exhibit very attractive intrinsic magnetic properties. We primarily focus on the magnetic anisotropic properties arising from the high spin-orbit coupling of the Pb atom. Decomposing the total anisotropy into intra- and interatomic contributions reveals a significant deviation from single ion anisotropy model with strong symmetric anisotropic pair interactions present. Furthermore, we consider magnetic properties of ternary Pb-based 3*d*-intermetallics which recently have been reported as stable or metastable. Giant magnetic anisotropy is found on Pb atoms in these systems. The origin of such strong anisotropy in $La_{18}Co_{28}Pb_3$ appears from two sources: spin-orbit and interelectronic Breit couplings. The significance of Breit interaction for magnetic anisotropy in bulk systems has not been reported previously. It is expected that Breit coupling induced anisotropy is dominating in magnetic Pb-based magnets with lower dimensionality including thin films.

**Introduction**

Theoretical search for better permanent magnets usually focuses on optimizing three intrinsic magnetic properties: magnetization, Curie temperature ($T_c$), and magnetic anisotropy (MA) in the applied temperature range 0-200°C [1]. Unfortunately, these properties have their optimum at different chemical and structural conditions. For example, Fe and/or Co bulk magnetic compounds and alloys can exhibit high magnetization and $T_c$, but their MA is usually very low except hcp Co [2]. Enhancement of MA in these materials can be achieved by switching to low dimensional or impurity-like structures where the effective electronic bandwidth is reduced [3]. However, reducing dimensionality significantly lowers the $T_c$. Such a problem is naturally resolved in intermetallics containing 3*d* and 4*f* elements. Thus, the best hard permanent magnets so far have been found among compounds containing rare earth (RE) atoms (responsible for the large atomic-like anisotropy) and significant amounts of Fe (Co) atoms (responsible for the large magnetization and $T_c$) [4-9]. Because RE materials can be rare and difficult to obtain, their replacement or optimization in hard magnets has been driving considerable fundamental and applied research in several directions for decades, as summarized in **Table 1**.

The first direction is to optimize existing Nd- and Sm-based hard magnets [10-12]. It includes both adding small amounts of light elements (for instance, Sm-Co-N magnets) and substituting Nd, Sm, and especially Dy with less expensive RE elements such as, for instance, La or Ce. This may not always improve the magnetic properties of these materials but can affect other needed properties, including performance at high temperatures.

The second direction is more traditional and relates to creating anisotropic Fe (Co) rich materials [12-23]. These materials are cheap, and creating a hard magnet of such type would affect the magnetism industry tremendously. While many promising systems have been found experimentally and/or predicted theoretically, a realistic replacement for RE-based magnet has yet to be developed.

The third direction of this search is related to compounds containing $3d$ elements and heavy elements with the spin-orbit coupling (SOC) even larger than that in RE elements [24-27]. These include such known magnets as FePt and MnBi, where the value of SOC on Pt or Bi atoms is very large (0.2-0.5 eV). Unfortunately, despite many years of research, no commercial hard magnet from this group of materials has been achieved.

In this paper, we extend the search along the third direction to systems containing Pb, an element with a very high SOC. It should be noted that bulk systems containing Pb and $3d$ elements are mostly unstable. Despite such bulk instability, there has been significant experimental research in magnetic thin films containing Pb atoms [28-38]. The role of Pb in magnetic properties has not been discussed much, but high MA was noticed [31-38]. Large MA can be produced by several relativistic mechanisms. While large SOC of Pb can be considered as the traditional relativistic mechanism due to its heavy mass, large MA in films suggests possible significance of interelectronic Breit coupling (BC) [39-40]. The role of BC in the bulk Pb-based system is unknown. In this paper, we first demonstrate that binary compounds of Pb and $3d$ transition metals can exhibit attractive magnetic properties due to the high SOC value of Pb atoms. We then discussed how to generate stable Pb-based magnets and analyze the origin of MA in the recently predicted La-Co-Pb and La-Fe-Pb ferromagnets. We show that Pb atoms play important roles in contributing to the MA in these systems, not only due to their high SOC but also by interelectronic BC. We emphasize that for the proper description of MA effects in Pb-based bulk and layered (films) magnets the inclusion of both SOC and BC is needed. Our results indicate that Pb-based intermetallic compounds could be promising candidates for applications in permanent magnetism and spintronics [41-44].

| Category | Materials | Notes |
| --- | --- | --- |
| $3d$-$4f$ intermetallics | CeCo$_5$, SmCo$_5$, Nd$_2$Fe$_{14}$B, Sm$_2$Fe$_{17}$N$_3$ | - LDA, LDA+U cannot describe crystal field splitting (CFS) in rare earth systems<br>- Need a fast and reliable method for $4f$ anisotropy<br>- Extensive experimental studies in the 1990s |
| Search for tetragonality in (Fe, Co)-rich systems | FeCo, FeNi, Fe$_{16}$N$_2$ | - LDA, LDA+U methods perform well<br>- Synthesis challenges |
| (Fe, Co)-rich systems with heavy elements (large SOC) | FePt, MnBi | - LDA, LDA+U methods perform well<br>- Synthesis challenges |
| | New Fe/Co+ heavy non-RE elements | - New findings in highly anisotropic structures for these combinations |

**Table 1.** Overview of the search for new permanent magnets.

**Computational Methods**

The first-principles calculations based on density functional theory (DFT) are performed by using the Perdew-Burke-Ernzerhof (PBE) exchange-correlation functional within the framework of GGA [45] as implemented in the Vienna Ab initio Simulation Package (VASP) [46-47]. A plane-wave cutoff energy is set to 520 eV. The accuracy of the electron self-consistent field is set to $10^{-4}$ eV, and the Brillouin zone is sampled using a set of gamma-centered uniform k-point sampling of $2\pi \times 0.025$ Å$^{-1}$. The lattice vectors of the unit cell and the positions of atoms in the unit are fully optimized with the force tolerance of 0.01 eV/ Å. The nonrelativistic spin-polarized calculations for collinear magnetism have been performed self-consistently [48]. Then a non-self-consistent calculation including SOC and based on the charge density of the previously self-consistent calculation is performed. When the SOC is included, symmetry operations are completely turned off and the spin-quantization axis is set to the chosen directions. For the MA calculations, we use a finer mesh size of $2\pi \times 0.016$ Å$^{-1}$ for k-point sampling to achieve better accuracy. For the MA decomposition calculations, we scale the SOC strength on different atoms with our implementation on a local version of VASP. BC has been included by using decomposition of the vector potential to intra- and interatomic contributions in full potential LAPW code developed by us previously [49]. Intraatomic contribution was treated according to the scheme proposed in [50], while interatomic interaction was included using the common far field approximation [51]. The energetic stability of the binary (ternary) compounds is evaluated by calculation the decomposition energy of the target binary (ternary) compound with respect to its nearby two (three) known stable phases on the convex hull, which we refer to as $E_{\text{hull}}$.

**Results and Discussions**

**A. Hypothetical binary Co$_3$Pb and Fe$_3$Pb compounds**

*Crystal structure and stability* -We begin by examining two hypothetical Pb containing binary magnetic compounds: Co$_3$Pb and Fe$_3$Pb. The theoretically proposed structures from OQMD database [52-53] for both compounds feature a hexagonal lattice within the *P6$_3$/mmc* space group, each unit cell comprising 8 atoms, as shown in **Fig. 1** (a). In this structure, Pb atoms bond with twelve equivalent Co (Fe) (1) atoms, forming a blend of distorted face and corner-sharing Pb-Co (Pb-Fe) cuboctahedra. Regarding magnetic symmetry, the 6 Co (Fe) atoms are segregated into two non-equivalent sites, with 4 atoms and 2 atoms, respectively. Our phonon calculations show that these two structures are dynamically stable without imaginary phonon modes as shown in **Fig. 1** (b) and (c).

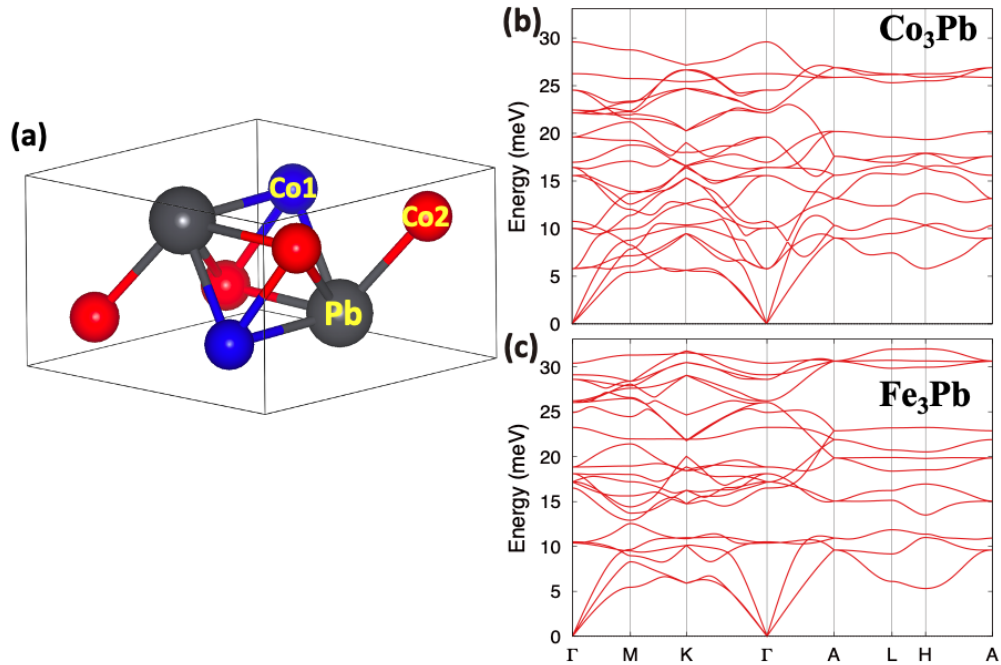

**Fig. 1** (a) The crystal structure of Co$_3$Pb. (b) and (c) The phonon band structure of Co$_3$Pb and Fe$_3$Pb.

*Electronic structures -* The electronic density-of-states (DOS) of Co$_3$Pb and Fe$_3$Pb are presented in **Fig. 2**. A prominent feature in Co$_3$Pb as shown in **Fig. 2** (b) is the high DOS near the Fermi level for the minority spin, primarily originating from Co atoms. This contrasting behavior between the up and down spin states is evident in the spin-polarized band structures displayed in **Fig. 2** (a) and (b). Notably, the DOS peak near the Fermi level for the minority spin is contributed from some flat bands, as depicted in **Fig. 2** (b), representing a Van Hove singularity.

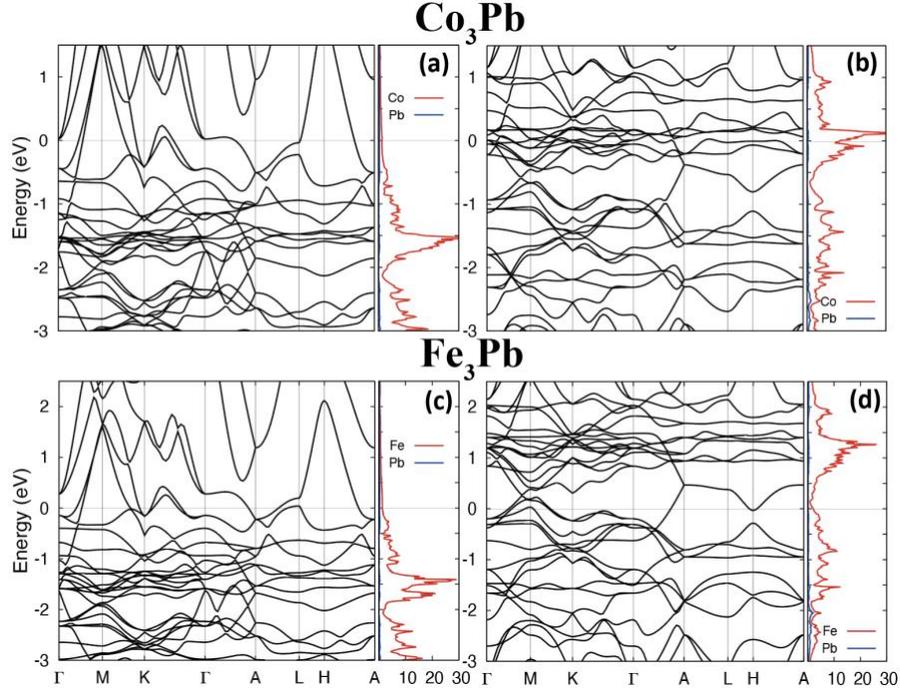

**Fig. 2** The electronic band structure with partial DOS of Co$_3$Pb with spin-up (a) and spin-down (b). The same for Fe$_3$Pb with spin-up (c) and spin-down (d) states.

In the case of Fe$_3$Pb, we observe a shift of the DOS peak for the minority spin towards the conduction states, as depicted in **Fig. 2** (d), attributed to the additional unpaired 3$d$ electrons in Fe. Fe$_3$Pb exhibits a magnetic moment of 2.44 $\mu_B$ for Fe which is higher than moment of bcc Fe and significantly larger than the 1.55 $\mu_B$ moment of Co atom obtained in Co$_3$Pb.

It is noteworthy that the Co $d$-states in the Co$_3$Pb compound manifest a relatively narrow bandwidth of approximately 3.5 eV, as can be seen from the DOS shown in **Fig. 2** (a) and (b). In elemental hcp Co, the band width of the $d$ states is about 4 - 4.5 eV. A smaller Fe $d$-band width is also observed for Fe$_3$Pb as compared to that in bcc Fe. These results suggest a potentially stronger electronic correlation in these binary compounds. We further investigate the sensitivity of the calculated electronic structure and magnetic properties of Co$_3$Pb and Fe$_3$Pb to possible variations of Coulomb correlations by conducting DFT+U calculations. Incorporating the U term on Co/Fe atoms in DFT-PBE calculations reveals a significant (~25% for Co and ~17% for Fe) and nearly linear increase of the magnetic moments of the Co/Fe atoms when the U value changes from 0 to 4 eV, as depicted in **Fig. 3** (a).

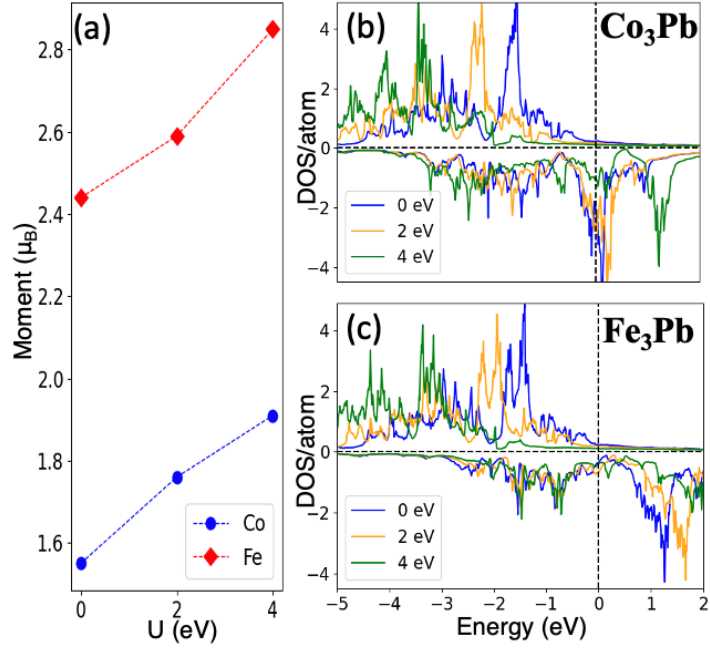

**Fig. 3.** (a) The calculated magnetic moments for Co/Fe sites in $Co_3Pb$ and $Fe_3Pb$ as a function of U on Co/Fe atoms. (b) The partial DOS of Co atoms in $Co_3Pb$ and (c) The partial DOS of Fe atoms in $Fe_3Pb$ with different U parameter.

We calculate the electronic DOS using different U values, as depicted in **Fig. 3** (b) and (c). At a U value of 4 eV, both $Co_3Pb$ and $Fe_3Pb$ exhibit a notable shift in DOS peaks, accompanied by a significant narrowing of bandwidth. To estimate the extent of electron correlation in these two systems, we can compare their electronic structure at different U values with that of MnBi [24], which is known to exhibit strong correlation as verified experimentally. For MnBi, while calculation results using a U value of 4 eV agree with experimental observations, the DOS of MnBi [24] showcases much narrower bandwidth compared to those of $Co_3Pb$ or $Fe_3Pb$. Thus, we infer that such strong correlation in MnBi might not be applicable in these systems, advocating for a U value of 2 eV or lower to aptly characterize their electronic and magnetic properties.

*Magnetic anisotropy -* Let us switch to MA calculations. MA studies are performed by direct calculation of the total energy with SOC included and with magnetic moments oriented along the (100), (010), and (001) directions. The direction associated with the lowest total energy is labeled as the magnetic "easy" direction. The MA energy is expressed as the energy difference between states with different magnetization directions, e.g., K = E[100] – E[001].

Our calculated total uniaxial K is 5.00 meV/cell (6.93 MJ/m$^3$) for $Co_3Pb$ and 0.70 meV/cell (0.94 MJ/m$^3$) for $Fe_3Pb$. Clearly, for both compounds, the MA is significantly larger than that of hcp Co (0.4 MJ/m$^3$), while for $Co_3Pb$ it is strong enough to be considered for applications.

Due to unknown extent of Coulomb interactions in these systems, we also perform the MA calculations for different c/a ratio and for different U values as shown in **Fig. 4**. The equilibrium c/a ratio is 0.742 for $Co_3Pb$, and 0.750 for $Fe_3Pb$, respectively. One can see the dependence of MA on the c/a ratio is very different in the two systems. For $Co_3Pb$, the MAs with U = 0 eV and

2 eV are similar in the whole c/a ratio range. At the equilibrium c/a ratio, Co$_3$Pb reaches total uniaxial K of 6.93 (5.75) MJ/m$^3$ at U = 0 (2) eV. At U = 4 eV, the MA of Co$_3$Pb is very sensitive to the c/a ratio and changes to negative MA value (in-plane anisotropy) at the equilibrium c/a ratio. For Fe$_3$Pb, the value of MA is very sensitive to the choice of U, although the variation trend of the MA with c/a ratio is similar. At the equilibrium c/a ratio, the total uniaxial MA of 0.93 (8.85) MJ/m$^3$ for U = 0 (2) eV. The increase of U from 0 to 2 eV enhances the MA by nearly an order of magnitude. The MA for U = 4 eV is in between U = 0 eV and 2 eV results. Clearly, for both compounds, the MA is significantly larger than that of hcp Co (0.4 MJ/m$^3$). Especially, the MAs at U = 2 eV for both Co$_3$Pb are Fe$_3$Pb are strong enough to be considered for applications.

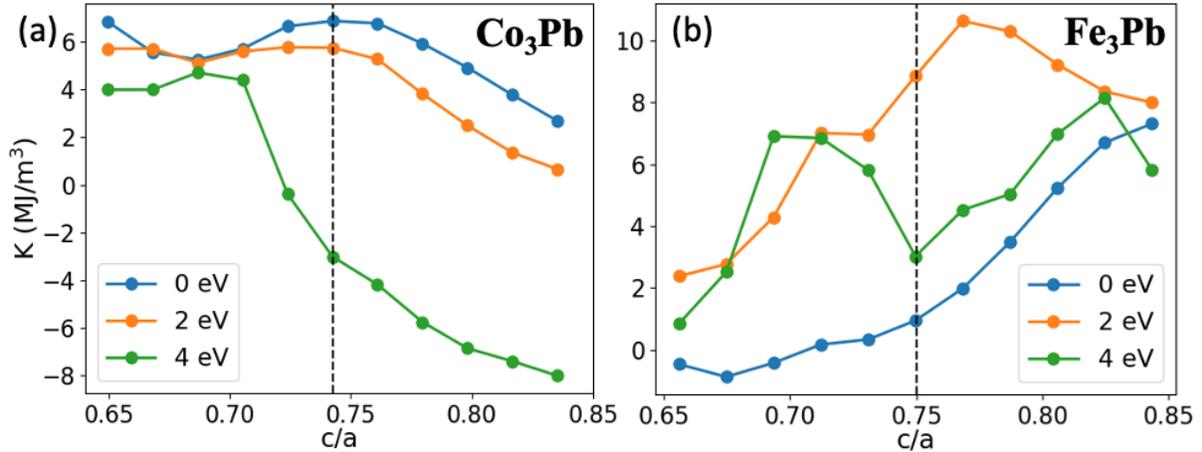

**Fig. 4.** The magnetic anisotropy energy with respect to different Hubbard U parameter and as a function of c/a ratio for Co$_3$Pb (a) and Fe$_3$Pb (b). The dashed line indicates the equilibrium c/a ratios of Co$_3$Pb and Fe$_3$Pb.

We proceed with the decomposition of the total MA into the contributions from different atomic sites to investigate the spatial distribution of the MA enhancement. Within second-order perturbation theory [54-55], the MA can be decomposed into contributions from individual sites in the unit cell. We adopt the spatial decomposition scheme of Antropov *et al* [54-55], where we express the MA as a half of the sum of the atomic SOC anisotropies over all atomic sites in the unit cell:

$$K = \frac{1}{2}\sum_i K^{SO}(i)\ ^{MA}$$

In this scheme, the site-resolved MA (i.e., K$_{SO}$(*i*)) is defined as the difference in the SOC energy between two magnetization orientations at the site *i*. Such a decomposition using SOC matrix elements can be also useful as it allows to make additional orbital and spin decompositions and relate it to the electronic structure.

Also, we can modify the strength of SOC on a single atomic site to explore how the effective magnetic anisotropy energy K$_{so}$ on other atoms changes with varying SOC strength. This adjustment is achieved by introducing a scaling factor λ, as proposed by [55]. Gradually varying λ from 0 to 1, we perform self-consistent calculations with SOC to obtain the corresponding total energy and $K_{so}$ as a function of λ. For Co₃Pb and Fe₃Pb, which feature one Pb site and two non-equivalent Co/Fe sites according to magnetic symmetry, we apply λ to each non-equivalent site (Pb, M1, M2, where M = Co, Fe). While the change in $K_{so}$ on site i where λ is applied results from both intra-atomic (proportional to $λ^2$) and intersite terms ($λ_iλ_j$), the change in $K_{so}$ on other sites arises solely from intersite interactions. Thus, variations in SOC on a given site reveal all pairwise interactions, and the functional dependence on λ allows determination of the effective spatial decomposition of anisotropy. In Co₃Pb and Fe₃Pb, we introduce two scaling parameters, $λ_1$ for the Pb atom and $λ_2$ for the Co/Fe atom, respectively. We compute all nearest pairwise interactions and formulate $K_{MA}$ for each non-equivalent site accordingly:

$$K_{MA-Pb} = λ_1^2 K_{Pb} + 6λ_1^2 K_{Pb-Pb} + 4λ_1λ_2 K_{Pb-M_1} + 8λ_1λ_2 K_{Pb-M_2}$$
$$K_{MA-M_1} = λ_2^2 K_{M_1} + 8λ_2^2 K_{M_1-M_2} + 4λ_1λ_2 K_{Pb-M_1}$$
$$K_{MA-M_2} = λ_2^2 K_{M_2} + 4λ_2^2 K_{M_2-M_2} + 4λ_2^2 K_{M_1-M_2} + 4λ_1λ_2 K_{Pb-M_2}$$

Where $K_X$ (X = Pb, M1, M2) are pure onsite contributions, and $K_{X-Y}$ (X, Y = Pb, M1, M2) are pairwise interactions.

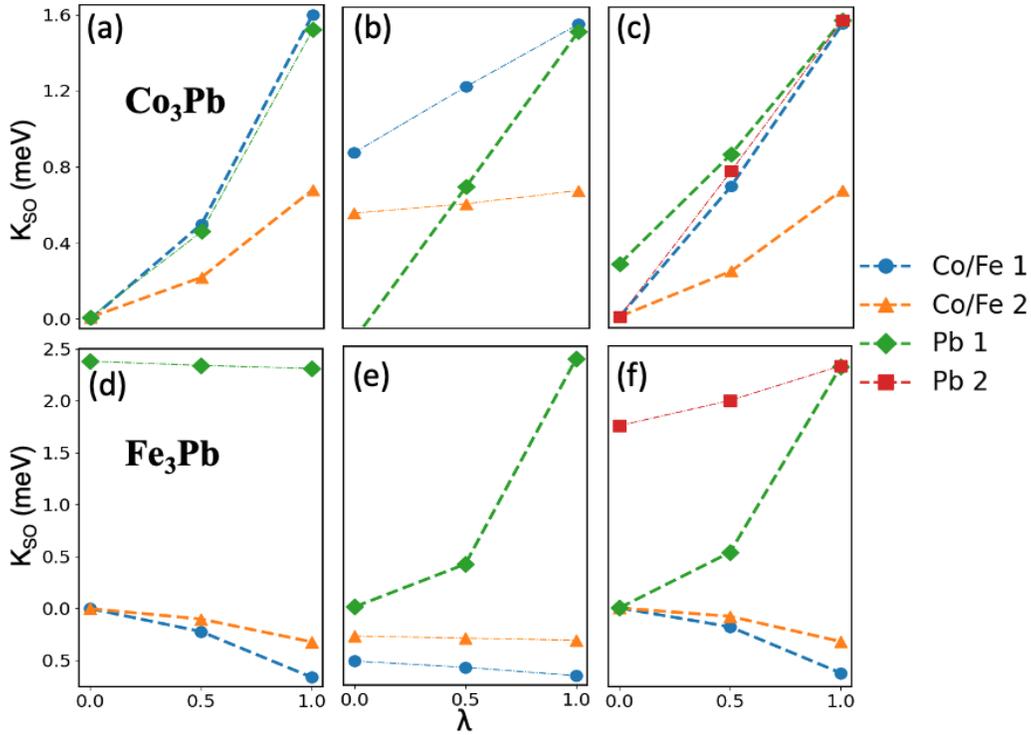

**Fig. 5.** The atomic SOC anisotropy energy $K_{SO}(i)$ as a function of the scaling factor λ (horizontal axis) on different non-equivalent site for Co₃Pb (top) and Fe₃Pb (bottom). Blue curves are Co/Fe 1 site, orange curves are Co/Fe 2 site, green and red curves are Pb atoms. The dashed bold line indicates where the scaling factor λ is applied on. λ is applied on all Co/Fe atoms in the left figures, on all Pb atoms in the middle figures, and on all Co/Fe atoms and one Pb atom in the right figures. The other atoms without scaling are shown as dash dot line with smaller linewidth. (U dependence plots are shown in the in Supplemental materials as **Fig. S1** and **Fig. S2**.)

To dissect the atomic magnetic anisotropies, we engage SOC of the valence electrons on specific atomic sites, as described earlier. **Fig. 5** illustrates the effective magnetic anisotropy energy $K_{so}$ on all non-equivalent sites in the unit cell as a function of the SOC scaling factor $\lambda$ on site X (X=Pb, M1, M2). In **Fig. 5** (a) and (d), the change in $K_{so}$ on all atoms is depicted for $\lambda$ applied across all Co/Fe sites. Notably, the anisotropies on Co/Fe sites manifest as linear functions, where Co sites contribute positively, and Fe sites contribute negatively to the MA energy. The behavior of Pb atoms differs significantly between the two compounds. In $Co_3Pb$, the anisotropy of Pb increases by approximately 0.6 meV as $\lambda$ varies from 0 to 1, signifying a robust pairwise interaction between Pb and Co. Conversely, in $Fe_3Pb$, the anisotropy remains relatively unchanged, suggesting a somewhat "single-ion" character of Pb. Adjusting the spin-orbit coupling strength on Pb atoms (**Fig. 5** (b) and (e)) reveals minimal impact on the anisotropy of the Co2 sites, with the primary pairwise interaction between Pb and Co originating from the Co1 sites. Similarly, the weak pairwise interaction between Pb and Fe is evident (**Fig. 5** (d) and (f)). To compute all onsite and pairwise interactions comprehensively, we also vary $\lambda$ across all Co/Fe sites and a single Pb atom. Through these calculations, we fit the three equations above, as listed in **Table 2**. Notably, a significant intra-atomic contribution of Pb ($K_{Pb}$) is observed in $Fe_3Pb$ (1.75 meV at U=0 eV), with ignorable contribution from Fe atoms, akin to the case of $SmCo_5$ [56]. On the other hand, a large intra-atomic contribution of Co1 sites ($K_{M1}$) is noted in $Co_3Pb$ (1.60 meV at U=0 eV), along with significant pairwise interactions between Pb and Co1 sites ($K_{Pb-M1}$) of 0.18 meV (with 4 Pb-Co1 pairs), elucidating the disparate behavior of magnetic anisotropy in the two compounds. With the inclusion of Hubbard U, the intra-atomic contribution of Pb is enhanced in both $Co_3Pb$ and $Fe_3Pb$, while the contributions from Co/Fe is reduced. However, the pairwise interaction ($K_{Pb-M1}$) decreases in $Co_3Pb$, while it is enhanced in $Fe_3Pb$. Importantly, it is worth noting that the MAE is not solely constrained by intra-atomic and nearest-neighbor couplings; contributions from more distant neighbors must also be considered. Such itinerant effects are expected to strongly influence the temperature dependence of magnetic anisotropy in these systems. Through our dissection of the total relativistic anisotropic energy of two Pb containing 3$d$-intermetalic compounds with uniaxial magnetic anisotropy into onsite and intersite pairwise interactions, we found significant contributions from intersite symmetric anisotropic interactions (Kitaev terms), indicating that the single-ion anisotropy approximation is inadequate for addressing such intermetallic compounds.

| $K_{so}$ contributions (meV/atom) | Intrasite | | | Intersite | | | |
|---|---|---|---|---|---|---|---|
| | $K_{Pb}$ | $K_{M1}$ | $K_{M2}$ | $K_{Pb-Pb}$ | $K_{Pb-M1}$ | $K_{Pb-M2}$ | $K_{M1-M2}$ |
| $Co_3Pb$ (U=0 eV) | -0.11 | 1.60 | 0.20 | 0.07 | 0.18 | 0.03 | -0.10 |
| $Co_3Pb$ (U=2 eV) | 0.33 | 1.20 | 0.15 | 0.11 | 0.14 | 0.04 | -0.06 |
| $Fe_3Pb$ (U=0 eV) | 1.75 | -0.20 | -0.20 | 0.10 | -0.04 | -0.01 | -0.04 |
| $Fe_3Pb$ (U=2 eV) | 2.77 | 0.00 | 0.00 | 0.12 | 0.15 | 0.08 | -0.02 |
| $Fe_3Pb$ (U=4 eV) | 1.90 | 0.02 | 0.02 | -0.02 | 0.10 | -0.01 | -0.01 |

**Table 2.** The fitted intrasite and intersite contributions to SOC energies for $Co_3Pb$ and $Fe_3Pb$, with different U values. The $K_{so}$ for $Co_3Pb$ with U=4 eV is not shown due to results cannot be fitted.

*Curie temperature* - To estimate the critical temperature of magnetic phase transition ($T_c$) we used the KKR version [57] of spin polarized RKKY approximation [58-60] implemented in a code developed in ref. [61]. This approximation is better suited for relatively localized magnetic moments systems (for $Fe_3Pb$ one can expect the error below 10%), while for more itinerant systems (like $Co_3Pb$) the error can be above 20% (see discussion in Ref. [62]). The KKR calculations are performed using muffin-tin potentials with radii of 2.31 a.u. for Co and Fe, and 2.95 a.u. for Pb. Self-consistency is achieved within energy difference of $10^{-6}$ a.u ($2.7 \times 10^{-5}$ eV) on a k-point sampling of 7x7x8. From the converged charge density, exchange interactions $J_{ij}$ are calculated on a 6x6x7 supercell, including 1426 pair interactions. The results shown in **Table 3**, indicate that exchange interactions in $Co_3Pb$ are relatively short ranged and first NN approximation would be appropriate while $Fe_3Pb$ exhibits more long-ranged behavior of the exchange interactions. Our calculations predict mean field $T_c$ to be 540 K for $Co_3Pb$ and 1387 K for $Fe_3Pb$. To assess the $T_c$ dependency on the value of U, we analyze the energy difference between ferromagnetic (FM) and antiferromagnetic (AFM) states at various U values, as shown in **Table 4**. Notably, the estimations at U = 0 eV are consistent with KKR results, showing a significantly larger energy difference between AFM and FM states in $Fe_3Pb$, indicative of a substantially higher $T_c$. For $Co_3Pb$, the energy difference increases notably with larger U values, while the increase is comparatively smaller for $Fe_3Pb$.

| Index | site1 | site2 | Distance (a) | $J_{ij}$ (meV) | multiplicity |
|---|---|---|---|---|---|
| 1 | Co2 | Co2 | 0.43 | 14.59 | 2 |
| 2 | Co2 | Co1 | 0.45 | 11.92 | 4 |
| 3 | Co2 | Co1 | 0.56 | 0.96 | 2 |
| 4 | Co2 | Co2 | 0.62 | -0.66 | 2 |
| 5 | Co2 | Co2 | 0.72 | 1.81 | 2 |
| 6 | Co2 | Co2 | 0.86 | -1.53 | 4 |
| 1 | Fe2 | Fe2 | 0.43 | 32.62 | 2 |
| 2 | Fe2 | Fe1 | 0.45 | 21.58 | 4 |
| 3 | Fe2 | Fe1 | 0.57 | 7.24 | 2 |
| 4 | Fe2 | Fe2 | 0.62 | 8.59 | 2 |
| 5 | Fe2 | Fe2 | 0.72 | -0.70 | 4 |
| 6 | Fe2 | Fe2 | 0.75 | -0.65 | 2 |
| 7 | Fe2 | Fe2 | 0.86 | -1.60 | 4 |
| 8 | Fe2 | Fe2 | 0.86 | 0.70 | 4 |
| 9 | Fe2 | Fe1 | 0.94 | 1.05 | 4 |
| 10 | Fe2 | Fe2 | 1.00 | 3.93 | 2 |
| 11 | Fe2 | Fe2 | 1.00 | -2.29 | 4 |

**Table 3.** The exchange coupling parameters $J_{ij}$ calculated with different inter-sites and distances in $Co_3Pb$ (top) and $Fe_3Pb$ (bottom). Only components larger than 0.5 meV are listed in the table. The distances of pair interactions are in unit of lattice constant a.

| $E_{AFM}-E_{FM}$ (meV/cell) | U=0 eV | U=2 eV | U=4 eV |
|---|---|---|---|
| $Co_3Pb$ | 0.56 | 0.89 | 1.79 |
| $Fe_3Pb$ | 1.60 | 1.92 | 1.94 |

**Table 4.** The energy differences between AFM and FM states for $Co_3Pb$ and $Fe_3Pb$, calculated with different U values.

With these calculations, in both cases, $T_c$ are predicted to be well above room temperature, indicating a potential usefulness of these compounds. The calculated formation energies of these two structures are 0.42 and 0.29 eV/atom respectively above the known convex hull, so it is challenging for bulk experimental synthesis. However, it is still possible to make them in thin films geometry [28-37], and we have shown that they are dynamically stable (see **Fig. 1**).

**B. Ternary La-Co-Pb compounds**

While $Co_3Pb$ and $Fe_3Pb$ display attractive magnetic moments and MA, they suffer from poor energetic stabilities owing to the immiscibility of the Co-Pb and Fe-Pb pairs. It has been demonstrated that incorporation of a suitable third element with immiscible pair of elements can form stable ternary compounds [63-64]. Several La-Co-Pb ternary compounds have been experimentally synthesized including $La_6Co_{13}Pb$, $La_5CoPb_3$, $La_{12}Co_6Pb$, and $La_4Co_4Pb$ phases [65-68]. Recently, we also predict a new $La_{18}Co_{28}Pb_3$ ternary compound using machine learning methods [69]. This compound exhibits favorable thermodynamic stability and magnetic properties with a formation energy only 1 meV/atom above the convex hull and an FM ground state with an average magnetic moment of 0.63 $\mu_B$ per Co atom [70]. Understanding the mechanism of MA in the ternary compound in comparison with that observed in the binary compounds discussed above would provide useful insights for better materials design and optimization for hard magnets. By substituting Co with Fe in the predicted $La_{18}Co_{28}Pb_3$ compound, we also obtain another metastable $La_{18}Fe_{28}Pb_3$ compound with a formation energy of 59 meV/atom above the convex hull (decomposed to $La+La_5Pb_3+Fe$). The $La_{18}Fe_{28}Pb_3$ compound has an FM ground state with an average magnetic moment of 1.55 $\mu_B$ per Fe atom.

The $La_{18}Co_{28}Pb_3$ and $La_{18}Fe_{28}Pb_3$ compounds have a tetragonal lattice with an I4/mmm space group symmetry with a complex unit cell containing 98 atoms, as shown in **Fig. 6** (a). There are four Wyckoff sites for La (4c,8i,16m,8j), three Wyckoff sites for Co/Fe (32o, 16n, 8h), and two for Pb (2b, 4e), respectively. Within the symmetrized unit cell, the atoms distribute several layers along the **c** direction. For each La layer, the La atoms form an inner ring with the center occupied by the Pb atom and four outer rings, which form multiple tetrahedra with Co/Fe o and n sites. These four outer rings are connected by an octahedron between each other with Co/Fe o and h sites. For $La_{18}Co_{28}Pb_3$, the lattice constant a = 14.02 Å and the c = 10.01 Å (c/a ratio of 0.714) is obtained after DFT-PBE optimization. The lattice constant a = 14.23 Å and the c = 10.02 Å is obtained for $La_{18}Fe_{28}Pb_3$. Both c/a ratios are close to $1/\sqrt{2}$.

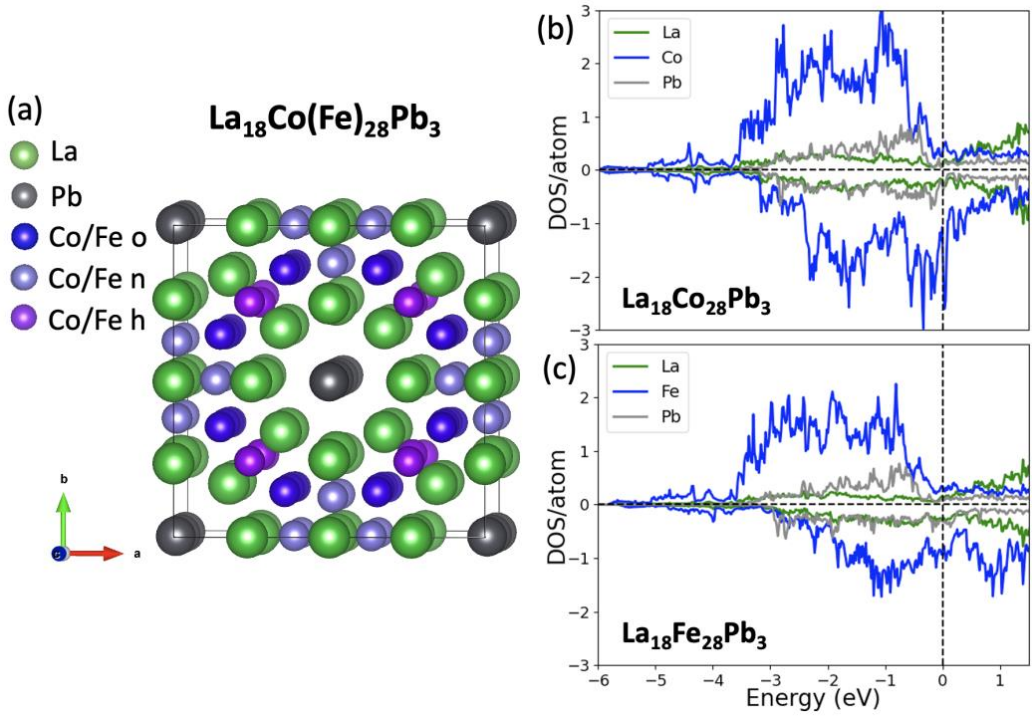

**Fig. 6.** (a) The crystal structure of the $La_{18}Co_{28}Pb_3$ ternary compound. The calculated DOS of the FM ground states of $La_{18}Co_{28}Pb_3$ (b) and $La_{18}Fe_{28}Pb_3$ (c).

Our calculations with SOC included revealed that the $La_{18}Co_{28}Pb_3$ has uniaxial MA of 11.1 meV/cell (1.06 MJ/m$^3$). Consecutive DFT+U calculations showed that with various U values, total MA remains substantially higher than that of hcp Co. In **Fig. 7**, we illustrate the total MA with each site's contributions as a function of U. Here, the significant MA predominantly arises from the Pb atoms, while the Co atoms show overall negative contributions. Particularly noteworthy is the substantial decrease observed on Pb e sites with the increasing U value.

We also calculate MA for the $La_{18}Fe_{28}Pb_3$ compound and found that it exhibits a significant strong in-plane MA magnetic anisotropy of -1.60 MJ/m$^3$, with an average anisotropy energy of -0.31 meV on each Fe atom. The anisotropy constant almost remains the same value when we change value of U from 0 to 4 eV.

Now we proceed with atomic decomposition of the MA in $La_{18}Co_{28}Pb_3$. **Fig. 7** (b) and (c) show the MA energy for all non-equivalent sites as a function of SOC scaling factor λ on Pb sites b and e. Notably, the MA on Co sites remain unaffected by variation of λ on any Pb site. Thus, unlike $Co_3Pb$ (see above) the pairwise SOC induced anisotropic interactions between Pb and Co atoms in this system are weak.

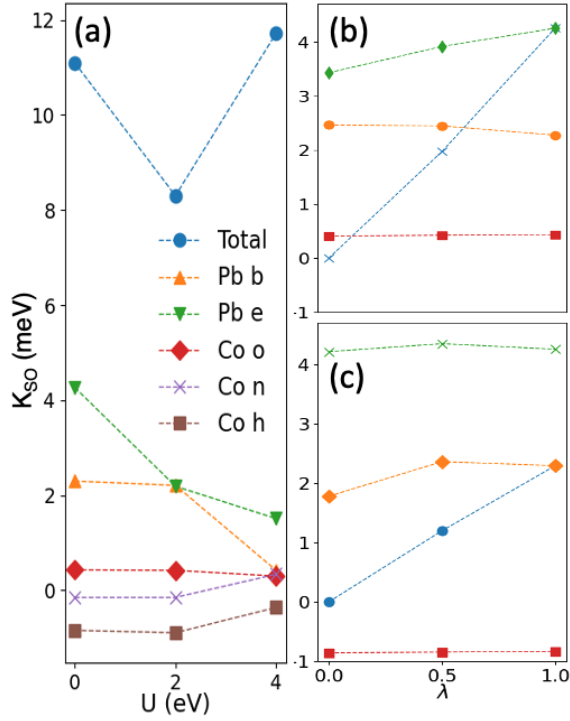

**Fig. 7.** (a) The calculated MA energies of $La_{18}Co_{28}Pb_3$ and their decomposition on each site as a function of U parameter on Co atoms. The atomic SOC anisotropy energy $K_{SO}(i)$ with U = 0 eV as a function of the scaling factor $\lambda$ on Pb b site (b) and Pb e site (c) for $La_{18}Co_{28}Pb_3$. The atomic SOC anisotropy energy $K_{SO}(i)$ with scaling factor applied is shown as blue curve. Other Pb b site is shown as orange curve, and Pb e site shown as green curve. Co n site (b) and h site (c) are shown in red curve, respectively, indicating no change in anisotropy energy with scaling factor on Pb atoms.

We also performed MA calculations including both the BC and the SOC interactions using full-potential LAPW method [49]. The results are shown in **Table 5**. In the case when only SOC was included this code produced results similar to that from VASP discussed above. Addition of BC leads to lowering of SOC anisotropy on Pb atoms but enhances the overall uniaxial MA by more than 60% in comparison to the result obtained only with SOC. The analysis of the different contribution to the MA from BC on Pb atoms indicates that the orbital term is dominating while spin term is relatively small. The contributions of Co atoms to the MA are also stronger and have primarily uniaxial character when both BC and SOC are included. While the total MA of Pb1 atom is two times smaller than anisotropy of Sm atom in $SmCo_5$, it is nearly four times larger than MA on Pt atom in FePt, highest known anisotropy due to SOC coupling for the bulk magnets. Such strength of MA is clearly competitive even with anisotropies of all known rare earth based anisotropic magnets.

| Atom | $K^*$ | $K_{SO}$ | $K_{BC1}$ | $K_{BC2}$ | $K_{total}$ |
|---|---|---|---|---|---|
| Pb1 | 5.14 | 4.64 | 2.97 | 1.23 | 8.84 |
| Pb2 | 3.22 | 2.86 | 1.74 | 0.62 | 5.22 |
| Co (total) | 0.12 | 0.17 | 0.06 | 0.02 | 0.25 |

**Table 5**. MA atomic decomposition of $La_{18}Co_{28}Pb_3$ in meV/atom. $K^*$ is atomic MAs with SOC only, $K_{SO}$ is the same as $K^*$ with BC included, $K_{BC1}$ is intraatomic BC contribution, $K_{BC2}$ is interatomic BC contribution. $K_{total}$ is total atomic MA by summing both SOC and BC contributions. BC was calculated with SOC included.

Overall, while the Pb intrasite contributions to MA are found significant, some notable intersite interactions of both SOC and BC nature between Pb atoms are also present. The large values of BC anisotropy already in this bulk Pb-based system suggest that the addition of BC terms is needed in all magnets with Pb atoms, and a domination of BC mechanism of MA in systems with lower dimensionality including layered (films) ones or having a specific shape could be also expected. Simultaneous inclusion of SOC and BC seems necessary as well.

**Conclusion**

We studied the magnetic properties of $3d$-based intermetallic systems containing Pb atoms. Our study demonstrates that while the abundance of $3d$ elements contributes to high Curie temperature and magnetization, the strong spin-orbit coupling of Pb produces a large magnetic anisotropy, both uniaxial and in-plane. For the binary $Co_3Pb$ phase, we found that this anisotropy is not of single-ion nature as seen in rare earth magnets, but likely has sizeable interatomic contributions due to strong d- and p-electron interactions. Although binary $Co_3Pb$ and $Fe_3Pb$ compounds are metastable in bulk form, we propose that thin films of these materials may stabilize and utilize their high magnetic anisotropy for practical applications.

We also explore the magnetic properties of recently predicted stable ternary Pb-magnets. We found a strong dependence of intrinsic magnetic properties on pressure and the strength of correlation effects in $La_{18}Co_{28}Pb_3$ and $La_{18}Fe_{28}Pb_3$. Our study emphasizes the significant role of Pb atoms in enhancing magnetic anisotropic properties. We demonstrated that an accurate description of magnetic anisotropy in $La_{18}Co_{28}Pb_3$ requires the inclusion of both SOC and Breit magnetic interelectronic couplings. Such importance of Breit interaction in bulk systems was not identified previously. The inclusion of Breit interaction is also expected to be crucial for magnetic anisotropy in Pb-based magnetic systems of low dimensions including films and surfaces, a topic we plan to explore in future studies. Simultaneous inclusion of SOC and BC is essential for a comprehensive understanding of anisotropic properties in these systems. The amplitude of obtained magnetic anisotropy on Pb atoms is competitive with magnetic anisotropies in the best rare earth-based magnets. Our results also suggest adding Breit coupling to standard electronic structures codes for the calculations of magnetic anisotropy especially for the systems with heavy atoms.

Our study broadens the search for novel magnetic materials to include Pb-based $3d$-intermetallics, with applications for permanent magnetism and spintronics areas. The insights gained from this work could pave the way for developing new magnetic materials with tailored properties for advanced technological applications. Future research should focus on the practical synthesis and experimental validation of these predicted compounds, particularly in thin-film forms, to fully realize their potential.

Acknowledgments

Work at Ames Laboratory was supported by the U.S. Department of Energy (DOE), Office of Science, Basic Energy Sciences, Materials Science and Engineering Division, including a grant of computer time at the National Energy Research Supercomputing Center (NERSC) in Berkeley. Ames Laboratory is operated for the U.S. DOE by Iowa State University under contract # DE-AC02-07CH11358.

# References

[1] J. M. D. Coey, Magnetism and Magnetic Materials. (Cambridge University Press, 2010).
[2] J. Fidler, Review of Bulk Permanent Magnets. In: G.C. Hadjipanayis, Magnetic Hysteresis in Novel Magnetic Materials. NATO ASI Series, vol 338. Springer, Dordrecht (1997).
[3] K.H.J. Buschow and F.R. Boer, Physics of magnetism and magnetic materials (Vol. 7, pp. 11-17). New York: Kluwer Academic/Plenum Publishers (2013).
[4] K. H. J. Buschow, P. A. Naastepad, and F. F. Westendorp, Preparation of SmCo5 permanent magnets, J. Appl. Phys. 40, 4029 (1969).
[5] D. Givord, H. S. Li, and J. M. Moreau, Magnetic properties and crystal structure of Nd2Fe14B, Solid State Commun. 50, 497 (1984).
[6] D. Givord, H. S. Li, and R. Perrier de la Bâthie, Magnetic properties of Y2Fe14B and Nd2Fe14B single crystals, Solid State Commun. 51, 857 (1984).
[7] J. M. D. Coey, H. Sun, Improved magnetic properties by treatment of iron-based rare earth intermetallic compounds in anmonia, J. Magn. Magn. Mater. 87, L251-L254 (1990).
[8] J. F. Herbst, R2Fe14B materials: Intrinsic properties and technological aspects. Rev. Mod. Phys. 63, 819-898 (1991).
[9] T. Miyake, K. Terakura, Y. Harashima, H. Kino, S. Ishibashi, First-Principles Study of Magnetocrystalline Anisotropy and Magnetization in NdFe12, NdFe11Ti, and NdFe11TiN. J. Phys. Soc. Jpn. 83, 043702 (2014).
[10] Y. Hirayama, T. Miyake, K. Hono, Rare-Earth Lean Hard Magnet Compound NdFe$_{12}$N. *JOM* **67**, 1344-1349 (2015).
[11] T. Miyake, H. Akai, Quantum Theory of Rare-Earth Magnets. J. Phys. Soc. Jpn. 87, 041009 (2018).
[12] J. M. D. Coey, Perspective and prospects for rare earth permanent magnets, Engineering 6, 119 (2020).
[13] M. D. Kuz'min, K. P. Skokov, H. Jian, I. Radulov, and O. Gutfleisch, Towards high-performance permanent magnets without rare earths, J. Phys.: Condens. Matter 26, 064205 (2014).
[14] H. Zhang, High-throughput design of magnetic materials. *Electronic Structure* **3**, 033001 (2021).
[15] G. A. Landrum, H. Genin, Application of machine-learning methods to solid-state chemistry: ferromagnetism in transition metal alloys. *Journal of Solid State Chemistry* **176**, 587-593 (2003).

[16] I. Miyazato, Y. Tanaka, K. Takahashi, Accelerating the discovery of hidden two-dimensional magnets using machine learning and first principle calculations. *Journal of Physics: Condensed Matter* **30**, 06LT01 (2018).

[17] S. Arapan, P. Nieves, S. Cuesta-López, A high-throughput exploration of magnetic materials by using structure predicting methods. *Journal of Applied Physics* **123**, 083904 (2018).

[18] T. Hasegawa, T. Niibori, Y. Takemasa, M. Oikawa, Stabilisation of tetragonal FeCo structure with high magnetic anisotropy by the addition of V and N elements. *Scientific Reports* **9**, 5248 (2019).

[19] Liqin Ke, Kirill D. Belashchenko, Mark van Schilfgaarde, Takao Kotani, and Vladimir P. Antropov, Effects of alloying and strain on the magnetic properties of $Fe_{16}N_2$, Phys. Rev. B 88, 024404 (2013).

[20] Xin Zhao, Cai-Zhuang Wang, Yongxin Yao, Kai-Ming Ho, Large magnetic anisotropy predicted for rare-earth free $Fe_{16-x}Co_xN_2$ alloys, Phys. Rev. B 94, 224424 (2016).

[21] M. Sakurai *et al.*, Discovering rare-earth-free magnetic materials through the development of a database. *Physical Review Materials* **4**, 114408 (2020).

[22] W. Xia, M. Sakurai, B. Balasubramanian, T. Liao, R. Wang, C. Zhang, H. Sun, K.-M. Ho, J. R. Chelikowsky, D. J. Sellmyer, C.-Z. Wang, Accelerating the discovery of novel magnetic materials using machine learning–guided adaptive feedback. *Proc. Natl. Acad. Sci. U.S.A* **119** (47) e2204485119 (2022).

[23] T. Liao, W. Xia, M. Sakurai, R. Wang, C. Zhang, H. Sun, K.-M. Ho, C.-Z. Wang, and J. R. Chelikowsky, Magnetic iron-cobalt silicides discovered using machine-learning Phys. Rev. Materials 7, 034410 (2023).

[24] V. P. Antropov, V. N. Antonov, L. V. Bekenov, A. Kutepov, and G. Kotliar, Phys. Magnetic anisotropic effects and electronic correlations in MnBi ferromagnet, Phys. Rev. B 90, 054404 (2014)

[25] V. N. Antonov and V. P. Antropov, Low-temperature MnBi alloys: Electronic and magnetic properties, constitution, morphology and fabrication (Review article), Low Temp. Phys. 46, 1 (2020)

[26] K. Son, G. H. Ryu, H.-H. Jeong, L. Fink, M. Merz, P. Nagel, S. Schuppler, G. Richter, E. Goering, G. Schütz, Superior Magnetic Performance in FePt L10 Nanomaterials. Small 15, 1902353 (2019).

[27] Gutfleisch, O., Lyubina, J., Müller, K.-.H. and Schultz, L. (2005), FePt Hard Magnets. Adv. Eng. Mater., 7: 208-212 (2005).

[28] M.M. Rahman, K.M. Krishna, T. Soga, T. Jimbo, M. Umeno, Optical properties and X-ray photoelectron spectroscopic study of pure and Pb-doped $TiO_2$ thin films Author links open overlay panel, Journal of Physics and Chemistry of Solids 60, 2, 201-210 (1999).

[29] A. Rajapitamahuni, L. L. Tao, Y. Hao, J. Song, X. Xu, E. Y. Tsymbal, and X. Hong, Ferroelectric polarization control of magnetic anisotropy in $PbZr_{0.2}Ti_{0.8}O_3/La_{0.8}Sr_{0.2}MnO_3$ heterostructures, Phys. Rev. Materials 3, 021401(R) (2019).

[30] D. Lükermann, M. Gauch, M. Czubanowski, H. Pfnür, and C. Tegenkamp, Magnetotransport in anisotropic Pb films and monolayers, Phys. Rev. B 81, 125429(2010).

[31] S. Peng, M. Wang, H. Yang et al, Origin of interfacial perpendicular magnetic anisotropy in MgO/CoFe/metallic capping layer structures, Sci Rep 5, 18173 (2015).

[32] K.-R. Hao, Y. Song and L. Zhang, Giant magnetic anisotropy of adatoms on the graphane surface, Nanoscale, 15, 11909-11 (2023).

[33] T. S. Plaskett, E. Klokholm, D. C. Cronemeyer, P. C. Yin, S. E. Blum Crossmark, Magnetic anisotropy in Pb‐substituted Eu3Fe5O12 films, Appl. Phys. Lett. 25, 357–359 (1974).

[34] D. -L. Sun, D. -Y. Wang, H. -F. Du, W. Ning, J. -H. Gao, Y. -P. Fang, X. -Q. Zhang, Y. Sun, Z. -H. Cheng, J. Shen, Uniaxial magnetic anisotropy of quasi-one-dimensional Fe chains on Pb/Si, Appl. Phys. Lett. 94, 012504 (2009).

[35] X. Ma and J. Hu, Large Perpendicular Magnetocrystalline Anisotropy at the Fe/Pb(001) Interface, ACS Appl. Mater. Interfaces 10, 15, 13181–13186 (2018).

[36] Z. Wang, Y. Yang, R. Viswan, J. Li, D. Viehland, Giant electric field controlled magnetic anisotropy in epitaxial BiFeO3-CoFe2O4 thin film heterostructures on single crystal Pb(Mg1/3Nb2/3)0.7Ti0.3O3 substrate, Appl. Phys. Lett. 99, 043110 (2011).

[37] W. Jahjah, J.-Ph. Jay, Y. Le Grand, A. Fessant, A.R.E. Prinsloo, C.J. Sheppard, D.T. Dekadjevi, and D. Spenato, Electrical Manipulation of Magnetic Anisotropy in a Fe81Ga19/Pb (Mg1/3Nb2/3) O3-Pb (Zr$x$Ti1−$x$) O3 Magnetoelectric Multiferroic Composite, Phys. Rev. Applied 13, 034015 (2020); Erratum Phys. Rev. Applied 18, 029901 (2022).

[38] P. C. Dorsey, S. B. Qadri, K. S. Grabowski, D. L. Knies, P. Lubitz, D. B. Chrisey, J. S. Horwitz, Epitaxial Pb–Fe–O film with large planar magnetic anisotropy on (0001) sapphire , Appl. Phys. Lett. 70, 1173–1175 (1997).

[39] V. B. Berestetskii, E. M. Lifshitz and L. P. Pitaevskii. Quantum electrodynamics (Course of theoretical physics. V.4), §83, Pergamon Press (1982).

[40] J.F. Jansen, Magnetic anisotropy in density-functional theory, Phys. Rev. B 38, 8022-8029 (1988).

[41] B. Tudu and A. Tiwari, Recent Developments in Perpendicular Magnetic Anisotropy Thin Films for Data Storage Applications, Vacuum 146, 329-341 (2017).

[42] A. Hirohat, K. Yamada, Y. Nakatan, I. -L. Prejbeanu, B. Diény, P. Pirro, B. Hillebrands, Review on spintronics: Principles and device applications, Journal of Magnetism and Magnetic Materials 509, 166711 (2020).

[43] M. Misiorny, M. Hell, and M. Wegewijs, Spintronic magnetic anisotropy, Nature Phys 9, 801–805 (2013).

[44] S. Bhatti, R. Sbiaa, A. Hirohata, H. Ohno, S. Fukami, S. N. Piramanayagam, Journal home page for Materials Today Research Review Spintronics based random access memory: a review, Materials Today 20, 9, 530-548 (2017).

[45] J.P. Perdew, K. Burke, K. and M. Ernzerhof, *Generalized Gradient Approximation Made Simple,* Phys. Rev. Lett. 77, 3865-3868 (1996).

[46] G. Kresse and J. Furthmüller, Efficiency of ab-initio total energy calculations for metals and semiconductors using a plane-wave basis set, Comput. Mater. Sci. 6, 15-50 (1996).

[47] G. Kresse and J. Furthmüller, Efficient iterative schemes for ab initio total-energy calculations using a plane-wave basis set. Phys. Rev. B 54, 11169-11186 (1996).

[48] S. Steiner, S. Khmelevskyi, M. Marsmann, G. Kresse, Calculation of the magnetic anisotropy with projected-augmented-wave methodology and the case study of disordered Fe$_{1-x}$Co$_x$ alloys. *Phys. Rev. B* **93**, 224425 (2016).

[49] N. E. Zein and V. P. Antropov, Self-Consistent Green Function Approach for Calculation of Electronic Structure in Transition Metals, Phys. Rev. Lett. 89, 126402 (2002).

[50] D. J. Griffiths, Introduction to Electrodynamics (5th ed.), Cambridge: Cambridge University Press (2023).

[51] A. L. Kutepov, Possible Way To Describe Breit's Interaction in Solids Composed From Heavy Elements, https://www.osti.gov/servlets/purl/950090, https://doi.org/10.2172/950090.

[52] Saal, J. E., Kirklin, S., Aykol, M., Meredig, B., and Wolverton, C. "Materials Design and Discovery with High-Throughput Density Functional Theory: The Open Quantum Materials Database (OQMD)", *JOM* **65**, 1501-1509 (2013).

[53] Kirklin, S., Saal, J.E., Meredig, B., Thompson, A., Doak, J.W., Aykol, M., Rühl, S. and Wolverton, C. "The Open Quantum Materials Database (OQMD): assessing the accuracy of DFT formation energies", *npj Computational Materials* **1**, 15010 (2015).

[54] Vladimir Antropov, Liqin Ke and Daniel Åberg, Constituents of magnetic anisotropy and a screening of spin-orbit coupling in solids, Solid State Commun. 194, 35 (2014).

[55] Yang Sun, Yong-Xin Yao, Manh Cuong Nguyen, Cai-Zhuang Wang, Kai-Ming Ho, and Vladimir Antropov, Spatial decomposition of magnetic anisotropy in magnets: Application to doped $Fe_{16}N_2$, Phys. Rev. B 102, 134429 (2020).

[56] Peter M. Oppeneer, Helmut Eschrig, and Börje Johansson, Calculated crystal-field parameters of $SmCo_5$ Manuel Richter, Phys. Rev. B 46, 13919 (1992).

[57] A.I. Liechtenstein, M.I. Katsnelson, V.P. Antropov, V.A. Gubanov, Local spin density functional approach to the theory of exchange interactions in ferromagnetic metals and alloys, Journal of Magnetism and Magnetic Materials, 67(1), 65-74 (1987).

[58] S. H. Liu, Quasispin model of itinerant magnetism: High-temperature theory, Phys. Rev. B 15, 4281 (1977).

[59] R. E. Prange and V. Korenman, Local-band theory of itinerant ferromagnetism. IV. Equivalent Heisenberg model, Phys. Rev. B 19, 4691 (1979).

[60] J. F. Cooke, Role of electron-electron interactions in the RKKY theory of magnetism, J. Appl. Phys. 50, 1782 (1979).

[61] N H Long and H Akai, First-principles KKR-CPA calculation of interactions between concentration fluctuations, *J. Phys.: Condens. Matter* **19** 365232 (2007).

[62] V. P. Antropov, The exchange coupling and spin waves in metallic magnets: removal of the long-wave approximation, J. of Magn. Magn. Mater. 262 (2), L192-L197 (2003).

[63] P. C. Canfield, S. L. Bud'Ko, FeAs-based super- conductivity: a case study of the effects of transion metal doping on $BaFe_2As_2$, Annu. Rev. Condens. Matter Phys. 1, 27–50 (2010).

[64] V. Taufour, U. S. Kaluarachchi, R. Khasanov, M. C. Nguyen, Z. Guguchia, P. K. Biswas, P. Bonfà, R. D. Renzi, X. Lin, S. K. Kim, E. D. Mun, H. Kim, Y. Furukawa, C.-Z. Wang, K.-M. Ho, S. L. Bud'ko, and P. C. Canfield, Ferromagnetic Quantum Critical Point Avoided by the Appearance of Another Magnetic Phase in $LaCrGe_3$ under Pressure, Phys. Rev. Lett. 117, 037207 (2016).

[65] A. M. Guloy, J. D. Corbett, Exploration of the interstitial derivatives of $La_5Pb_3$ ($Mn_3Si_3$-type), Journal of Solid State Chemistry 109, 352–358 (1994).

[66] F. Weitzer, A. Leithe-Jasper, P. Rogl, K. Hiebl, H. Noël, G. Wiesinger, W. Steiner, Magnetism of (Fe, Co)-based alloys with the $La_6Co_{11}Ga_3$-type. Journal of Solid State Chemistry 104, 368–376 (1993).

[67] L. Gulay, Y. M. Kalychak, M. Wolcyrz, K. Lukaszewicz, Crystal structure of $R_{12}Ni_6Pb$ (R = Y, La, Pr, Nd, Sm, Gd, Tb, Dy, Ho) and $R_{12}Co_6Pb$ (R = Y, La, Pr, Nd, Sm, Gd) compounds, Journal of alloys and compounds 311, 238–240 (2000).

[68] T. J. Slade, N. Furukawa, M. Dygert, S. Mohamed, A. Das, W. Xia, C.-Z. Wang, S. L. Bud'ko, and P. C. Canfield, $La_4Co_4X$ (*X*=Pb, Bi, Sb): A demonstration of antagonistic pairs as a route to quasi-low-dimensional ternary compounds, Phys. Rev. Materials 8, 064401 (2024).

[69] R. Wang, W. Xia, X. Fan, H. Dong, T. J. Slade, K M Ho, P. C. Canfield, C-Z. Wang, ML-guided discovery of La-Co-Pb ternary compounds involving immiscible pairs of Co and Pb elements, npj Computational Materials 8 (1), 258 (2023).

[70] W. Xia, V. Antropov, Y. Yao, C. Z. Wang, Coexistence of low and high spin states in $La_{18}Co_{28}Pb$, Phys. Rev. B 109, 214429 (2024).

## Supplementary material

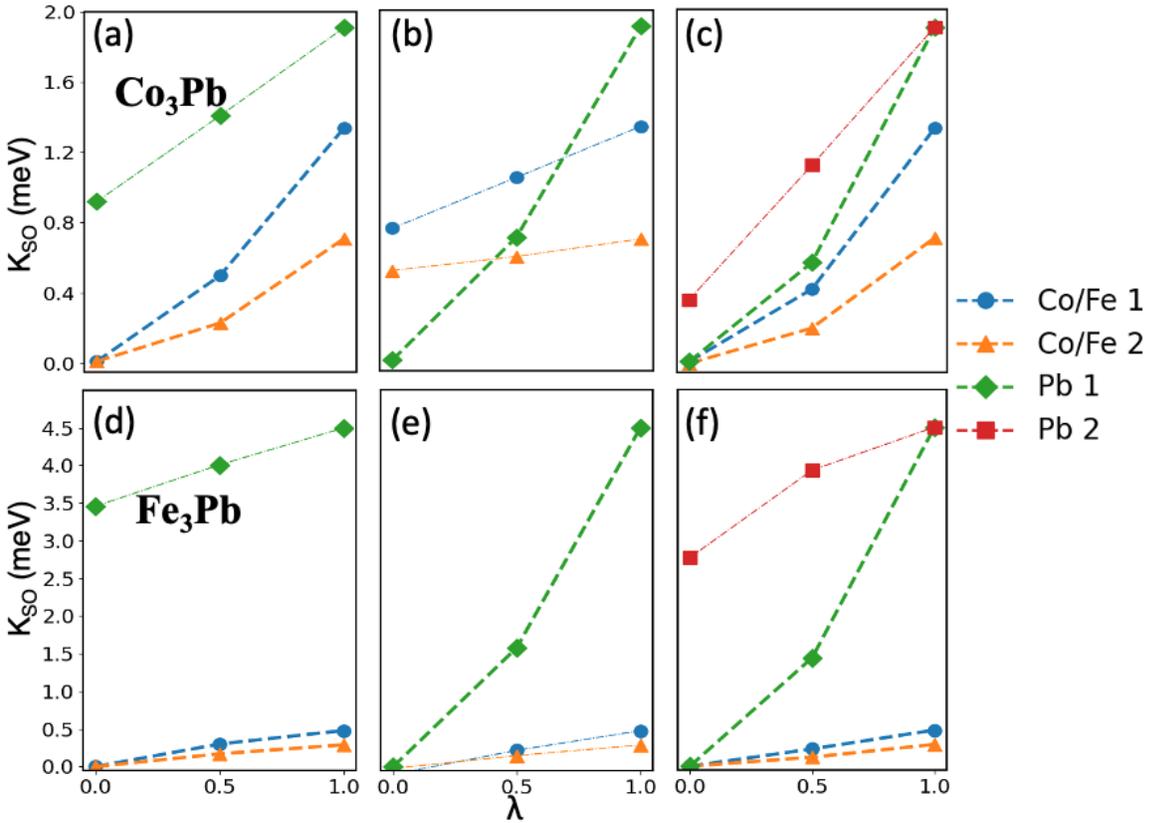

**Fig. S1.** The atomic SOC anisotropy energy $K_{SO}(i)$ as a function of the scaling factor $\lambda$ (horizontal axis) on different non-equivalent site for $Co_3Pb$ (top) and $Fe_3Pb$ (bottom) for $U = 2$ eV. Blue curves are Co/Fe 1 site, orange curves are Co/Fe 2 site, green and red curves are Pb atoms. The dashed bold line indicates where the scaling factor $\lambda$ is applied on. $\lambda$ is applied on all Co/Fe atoms in the left figures, on all Pb atoms in the middle figures, and on all Co/Fe atoms and one Pb atom in the right figures. The other atoms without scaling are shown as dash dot line with smaller linewidth.

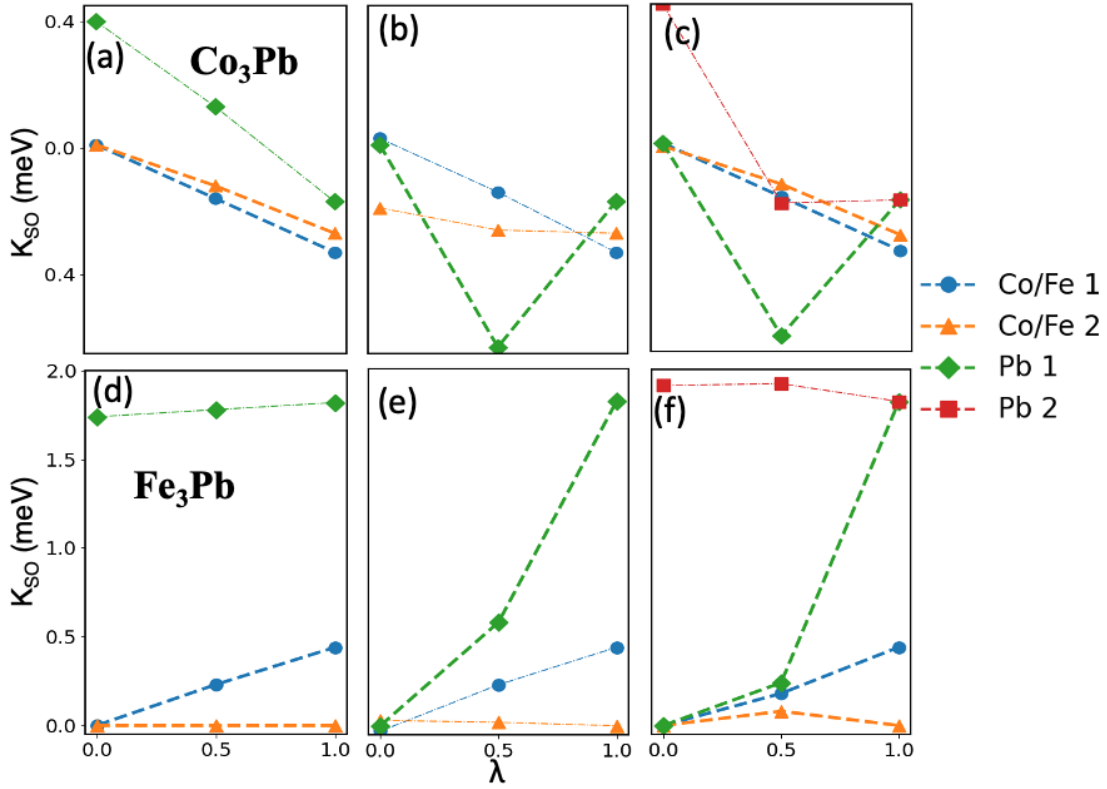

**Fig. S2.** The atomic SOC anisotropy energy $K_{SO}(i)$ as a function of the scaling factor $\lambda$ (horizontal axis) on different non-equivalent site for $Co_3Pb$ (top) and $Fe_3Pb$ (bottom) for $U = 4$ eV. Blue curves are Co/Fe 1 site, orange curves are Co/Fe 2 site, green and red curves are Pb atoms. The dashed bold line indicates where the scaling factor $\lambda$ is applied on. $\lambda$ is applied on all Co/Fe atoms in the left figures, on all Pb atoms in the middle figures, and on all Co/Fe atoms and one Pb atom in the right figures. The other atoms without scaling are shown as dash dot line with smaller linewidth.